\documentclass[english,twocolumn]{revtex4}
\usepackage[T1]{fontenc}
\usepackage[latin9]{inputenc}
\usepackage{graphicx}
\usepackage{amssymb}
\usepackage{amsmath}

\newcommand{\bea}{\begin{eqnarray}}

\newcommand{\eea}{\end{eqnarray}}

\newcommand{\beq}{\begin{equation}}

\newcommand{\eeq}{\end{equation}}



\usepackage{babel}

\begin{document}

\title{Editorial: Emergence in driven solid-state and cold-atom systems}

\author{Ludwig Mathey and Junichi Okamoto}

\affiliation{Zentrum f\"ur Optische Quantentechnologien and Institut f\"ur Laserphysik, Universit\"at Hamburg, 22761 Hamburg, Germany\\
The Hamburg Centre for Ultrafast Imaging, Luruper Chaussee 149, Hamburg 22761, Germany}

\begin{abstract}

\end{abstract}
\maketitle
While phase transitions in equilibrium systems have been studied extensively, the emergence of order in non-equilibrium
systems, such as periodically driven systems, continues to pose conceptual questions as well as offer intriguing
possibilities. 
In particular, two recent experimental developments emphasize the urgency of further exploration.
A new direction of research is taking place in the field of ultra-cold atoms.
 In a recent set of experiments, atoms in an optical lattice were subjected to a lattice shaking protocol.
   With this new feature of ultra-cold atom systems, it has been possible to create effective, renormalized Hamiltonians. In particular, the lattice shaking could be performed in such a manner to create frustrated systems and synthetic gauge fields, see e.g. Ref. \cite{struck}. 
A parallel development is taking place in solid state physics. A new research direction was established by using  ultrafast light pulses
 to stabilize superconducting order. Here, the periodic driving was used to achieve a remarkable, counterintuitive
result: to enhance superconductivity with high frequency and high intensity driving, see e.g. Ref. \cite{hu}. 



In the Focus Section compiled in this issue, the background and the conceptual environment of these studies are explored and developed. The contributions by Heinisch and Holthaus \cite{Heinisch} and the one by Eckardt and Str\"{a}ter \cite{Eckardt} address the crucial question of how large the heating rate and the entropy production of periodic driving is. This question has to be addressed because the heating effects are a key limitation to the control of an isolated system, such as cold atom systems,  via periodic driving. An insightful discussion of a periodically driven p-wave superconductor is given in the contribution by Zhao \cite{Zhao}. This discussion is important because the author takes a step beyond s-wave superconductors, which might be the simplest form of a superconducting state that comes to mind. Indeed, the high-T$_{c}$ superconductors that were used in Ref. \cite{hu} were not of s-wave symmetry. In the contribution by Mintert, Verdeny and Puig \cite{Mintert}, an interesting scenario is explored, in which the driving term is not purely periodic. Obviously, neither in cold atom or solid state systems, the driving term has to have a purely periodic form. So the article discusses a first non-trivial example of these vast possibilities.
 Finally, in the contribution by Zhu, Rexin and Mathey \cite{Zhu}, both the technical aspect of this field, how to generate the effective low-energy Hamiltonian, and the physically important question of how to stabilize the  dispersion of a system parametrically in an open environment is addressed.
 
 As can be seen from this selection of articles, the field of emergence in driven many-body systems is incredibly rich. Both the choice of equilibrium system and the choice of the  driving term, and what observable is to be controlled, allows for near-endless combinations. In the field of solid state physics especially, these ideas open a new direction of material design. While methods such as doping, using various material compounds, using different pressures and temperatures, are used routinely to influence the properties of a material, now dynamic control has been conceived as a new method.    We note that cold atom systems once again can serve as well-defined model systems, to understand emergence in driven systems systematically.
 
 We are merely starting to understand these exciting possibilities. We hope that this special issue will contribute to the development of this field, and inspire further study.

\begin{acknowledgments}
This work was supported by the Deutsche Forschungsgemeinschaft
through the SFB 925 and the Hamburg
Centre for Ultrafast Imaging, and from the Landesexzellenzinitiative
Hamburg, supported by the Joachim Herz Stiftung.

\end{acknowledgments}

\end{document}